\documentclass{iopjournal}

\usepackage[utf8]{inputenc}
\usepackage{amsmath}
\usepackage{booktabs}
\usepackage{url}
\graphicspath{{./}{figures/}}

\newcommand{\fl}{\,foot lengths}
\newcommand{\Dlim}{D_\mathrm{limit}}
\newcommand{\Lth}{L_\mathrm{th}}
\newcommand{\dcom}{d_\mathrm{COM}}

\renewcommand{\articletype}[1]{{\vspace*{-8mm}\noindent\small\sf Preprint submitted to European Journal of Physics}\par\vspace*{6mm}}
\begin{document}

\articletype{Paper}

\title{The chair challenge: can mechanics really tell men from women? A centre-of-mass analysis of a viral party trick}

\author{Ronald M Aarts$^{1,*}$\orcid{0000-0003-3194-0700} and Kees Schouhamer Immink$^{2}$\orcid{0000-0001-6747-9261}}

\affil{$^1$Department of Electrical Engineering (EE-SPS), Eindhoven University of Technology, Eindhoven, The Netherlands}

\affil{$^2$Turing Machines Inc, Rotterdam, The Netherlands}

\affil{$^*$Author to whom any correspondence should be addressed.}

\email{r.m.aarts@tue.nl, immink@turing-machines.com}


\begin{abstract}
In the ``chair challenge'' a person stands several foot lengths from a wall, bends forward until the head touches the wall, lifts a chair and tries to stand up. Online videos suggest that women can do this and men cannot, supposedly because women have a lower centre of mass or men have larger feet. We model the body as a planar chain of rigid segments with standard biomechanical parameters and search all postures allowed by the rules for the one that places the combined centre of mass of body and chair furthest back. As a rule of thumb, the largest toe--wall distance is roughly the trunk-plus-head length minus the distance from hip to centre of mass. For a slow, static rise this limit is 2.36\fl{} for the male and 2.49\fl{} for the female reference model, just short of the customary 2.5\fl. The limit is set mainly by trunk-plus-head length relative to foot length, not by the height of the centre of mass; in centimetres the male model actually has more room. With the body proportions of 6068 participants in the 2012 US Army anthropometric survey, the sex difference shrinks from 0.13 to 0.03--0.05\fl, half to one European shoe size. Individually, only 13\% of men and 22\% of women would pass statically, and predicting sex from the outcome is right in 54--68\% of cases (for equal numbers of men and women), against 50\% for a coin and 83\% for a tape measure. Shoes, age and momentum matter more than the residual sex difference. The problem suits first-year undergraduate mechanics, combining static equilibrium of an articulated body, scaling and units, an inverted-pendulum estimate and a simple classification problem. The chair challenge is a fine demonstration of static equilibrium and a lousy sex predictor.
\end{abstract}

\section{Introduction}
Few physics demonstrations have been filmed as often as the \emph{chair challenge}. The instructions are simple: stand facing a wall, step back two or three heel-to-toe foot lengths, bend at the hips until the top of the head rests against the wall, pick up a chair, lift it to the chest and stand up straight without moving the feet or using the hands against the wall~\cite{fatherly,athletico,kym}. The trick is older than the internet. It appears in the episode ``Judging Books by Covers'' of the CBS sitcom \emph{All in the Family}, first broadcast on 9 February 1971, in which daughter Gloria presents it as something women can do and men cannot, to the evident displeasure of her father Archie Bunker~\cite{kym,aitf,aitfclip}. A ``chair trick'' video appeared on YouTube in May 2009, and in mid-November 2019 the challenge went viral on TikTok, with single attempts collecting millions of views within days. Worldwide Google searches for it rose more than fifty-fold, and news coverage on three continents followed~\cite{kym,cbsny,gulf,capital}.

Almost all of this coverage frames the trick as a sex test and offers one of two explanations. The first is that women have a lower centre of mass, located near the hips rather than the chest~\cite{gulf,cbsny}. The second is that men have larger feet, so that stepping back a fixed number of foot lengths places them further from the wall~\cite{gulf,capital,fatherly,athletico}. In a widely viewed video a father who fails at the standard distance succeeds when he stands exactly where his wife stood~\cite{mathdad}. As far as we know, neither explanation has been tested quantitatively.

The question is a good one for a physicist: it reduces a social-media claim to a problem in static equilibrium, while the answer depends on human anatomy in an interesting way. Anyone who has tried the trick will notice that the body is not a rigid block. By tilting the ankles, bending the knees, arching the neck or holding the chair closer, one can move the combined centre of mass. The relevant question is therefore not where the centre of mass \emph{is}, but where it \emph{can be}. In this paper we compute the set of attainable centres of mass, find the most favourable one, and ask how much of the supposed sex difference survives. We did not ask anyone to lift a chair; all participants in this study are either simulated or were measured by the US Army for other purposes.

The paper is aimed at teachers and students of first-year undergraduate mechanics. Textbooks usually treat the centre of mass of rigid bodies; the chair challenge shows what changes when the body is articulated and constrained, and why a single sentence (``women have a lower centre of mass'') can be both true and irrelevant. Section~\ref{sec:teaching} suggests how the problem can be used in class.

\section{The rules}\label{sec:rules}
We formalise the challenge with six rules. R1: the tips of the toes are at a fixed horizontal distance from the wall, called the \emph{toe--wall distance} $D$, customarily 2.5 of the person's own foot lengths. R2: both feet stay flat and in place. R3: the head rests against the wall until standing up and may push against it. R4: the hands hold the chair and touch neither wall, floor nor legs. R5: the chair is lifted off the floor and held until upright. R6 (our addition, see section~\ref{sec:shoes}): barefoot or thin socks, $D$ paced in bare foot lengths. A static attempt succeeds if the combined centre of mass of body and chair can be brought directly above the feet, i.e.\ between heel and toe tip.

\section{Model}\label{sec:model}
\subsection{Body and chair}
The body is a sagittal chain of rigid segments: foot, shank, thigh, trunk, head, upper arm and forearm with hand, with both legs and both arms moving together. The stick figures in figure~\ref{fig:posture} show this chain. Segment lengths, masses and centre-of-mass locations are taken from De Leva's adjustment of the Zatsiorsky--Seluyanov parameters, which gives separate values for men and women based on gamma-ray scans of 100 male and 15 female subjects, mostly physical-education undergraduates~\cite{deleva}; these in turn build on the classical cadaver and living-subject data~\cite{dempster,clauser,winter}.

Throughout this paper \emph{stature}, symbol $H$, means standing body height, measured from the floor to the top of the head; we use ``stature'' rather than ``height'' to avoid confusion with the height of the centre of mass above the floor. The male reference model has $H=1.81$~m and a body mass of 84~kg, the female reference model $H=1.67$~m and 72~kg, typical adult values~\cite{cbs}. The segment lengths of De Leva are scaled with a single factor such that ankle height plus shank, thigh, trunk and head add up to $H$. Stature and mass only set the overall scale (sections~\ref{sec:sens} and~\ref{sec:shoes}); the body \emph{proportions}, which determine the result, come from Russian physical-education students (De Leva) and, as a check, from US Army personnel (section~\ref{sec:ansur}). Foot length $L_\mathrm{foot}$ is 15.2\% (man) and 15.0\% (woman) of stature, i.e.\ 27.5 and 25.1~cm. The chair has a mass of 4~kg, concentrated at the hands.

\subsection{Search over postures}\label{sec:search}
A posture is described by five variables: shank angle ($-10^\circ$ to $+25^\circ$ from vertical), knee flexion (0--90$^\circ$), trunk angle (40--140$^\circ$ from vertical), and the position of the grip along and below the trunk. Hip flexion is limited to $95^\circ$ plus half the knee flexion, representing hamstring tightness, and neck flexion/extension to $-45^\circ$/$+60^\circ$. For a given toe--wall distance $D$ the head angle follows from the requirement that the vertex touches the wall (R3); the elbow follows from inverse kinematics of the two-link arm (R4). Postures in which the chair passes through the legs, the floor or the wall are rejected.

For each value of the toe--wall distance $D$ we draw 150\,000 random postures per head solution (head tilted up or down), keep those that satisfy all constraints, compute the combined centre of mass, and refine the most rearward one by a local random search. The body itself is not varied: each search uses one fixed set of segment dimensions and masses, and only the joint angles and the grip position change. The \emph{geometric limit} $\Dlim$ is the largest toe--wall distance at which the best attainable centre of mass still lies above the tip of the toes; it is found by bisection on $D$. Repeated runs differ by less than 0.01\fl. The model code (Python) is available as described at the end of the paper.

\section{Results}
\subsection{The attainable centre of mass}
\begin{figure}[t]
\centering\includegraphics[width=\textwidth]{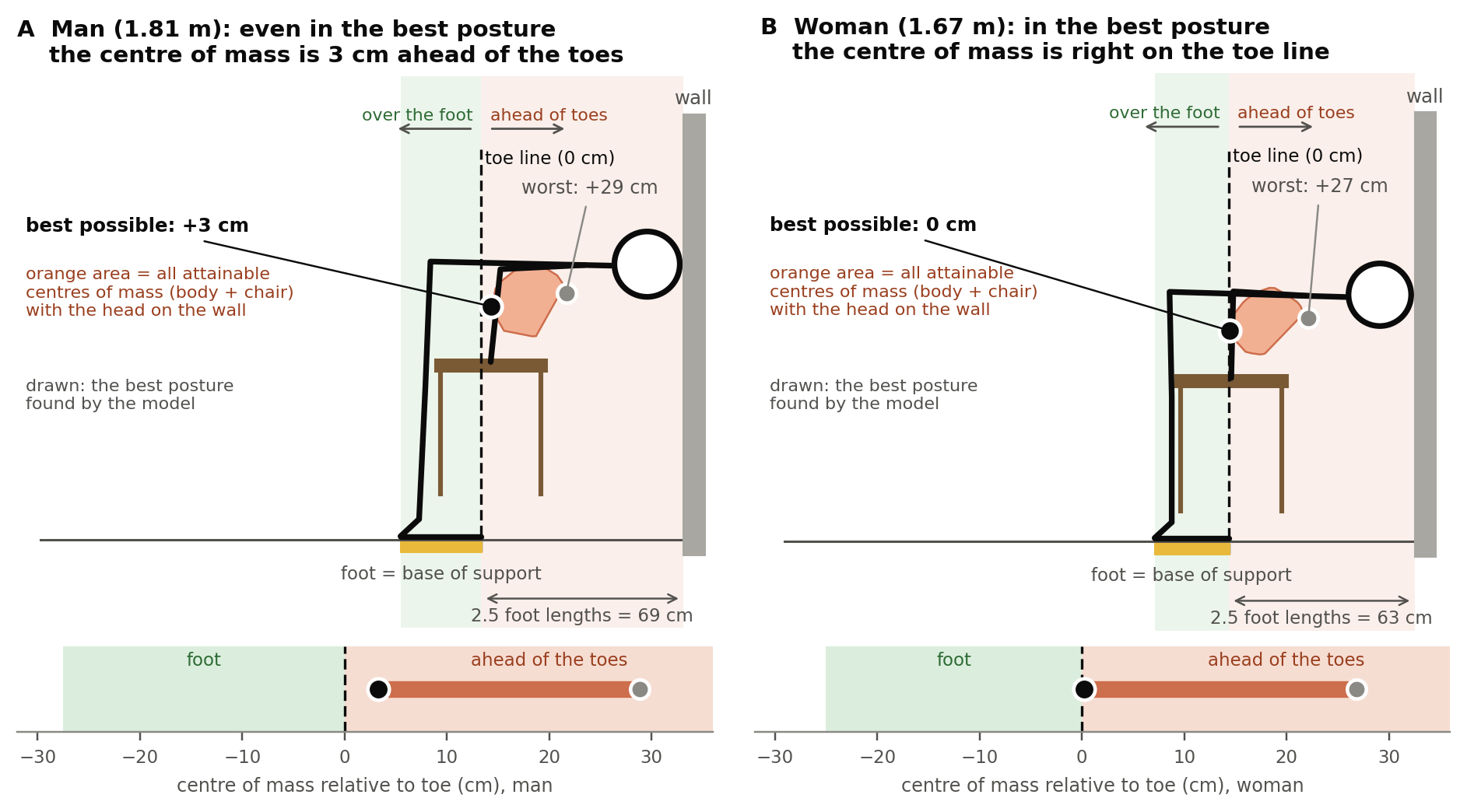}
\caption{Male (A) and female (B) reference model with the toes 2.5 of their own foot lengths from the wall, segment proportions of De Leva. The orange area contains all centres of mass of body plus chair compatible with the rules; the stick figure is the best posture found. The ruler below compares the attainable range with the foot (green) and the forbidden zone ahead of the toes (red).}
\label{fig:posture}
\end{figure}

Figure~\ref{fig:posture} shows the result for the customary toe--wall distance of 2.5\fl. The set of attainable centres of mass is small and lies mostly in front of the toes: from $+3$ to $+29$~cm for the man and from 0 to $+27$~cm for the woman. The best posture is always the same: legs straight and vertical, trunk horizontal, head in line with the trunk, and the chair held close under the abdomen, above the feet. Even in this posture the man's centre of mass is 3~cm ahead of his toes, so he cannot rise without momentum. The woman's centre of mass is right on the toe line.

\subsection{A rule of thumb}\label{sec:rule}
Why this posture, and what sets the limit? Take the hip joint as reference point. With the trunk horizontal and the head in line, the top of the head lies one trunk-plus-head length in front of the hip. Because the head must touch the wall, the hip is then exactly this length from the wall; a more upright trunk or a bent neck would bring the hip closer to the wall. In this posture the combined centre of mass of body and chair lies a horizontal distance $\dcom$ in front of the hip joint, and hence a distance $\Lth-\dcom$ from the wall. A static rise requires this centre of mass to lie above the feet, so the toes may be at most this far from the wall. This gives the rule of thumb
\begin{equation}
  \Dlim \approx \Lth - \dcom , \label{eq:rule}
\end{equation}
where
\begin{itemize}
\item $\Dlim$ is the geometric limit: the largest toe--wall distance at which a static rise is possible;
\item $\Lth$ is the trunk-plus-head length, the distance from the hip joint centre to the top of the head (vertex) with trunk and head in line;
\item $\dcom$ is the horizontal distance by which the combined centre of mass of body and chair lies in front of the hip joint in the best posture.
\end{itemize}
In figure~\ref{fig:posture} the hip is the corner between the legs and the horizontal trunk, $\Lth$ is the distance from that corner to the wall, and the black dot marks the combined centre of mass. With the scaled De Leva segments, $\Lth$ is 47.6\% of stature for the male and 49.5\% for the female reference model, i.e.\ $\Lth=86.1$~cm and 82.7~cm. Summing the moments of all segments and of the chair about the hip gives $\dcom=21.1$~cm (11.7\% of stature) for the man and 20.0~cm (12.0\%) for the woman. Equation~(\ref{eq:rule}) then gives $\Dlim\approx 86.1-21.1=65.0$~cm, or 2.36\fl, for the man and $82.7-20.0=62.7$~cm, or 2.50\fl, for the woman. The full model, i.e.\ the posture search of section~\ref{sec:search}, which also lets the ankles, knees and neck move, gives 65~cm (2.36\fl) and 62~cm (2.49\fl), within 0.01\fl{} of the rule of thumb.

To apply the rule to individuals we need $\dcom$ for other body proportions. Running the full model with $\Lth$ varied by $\pm3\%$ at fixed stature gives, to good approximation, $\dcom\approx 0.29\,\Lth - c\,H$, where $H$ is stature and $c$ is a dimensionless fit constant, $c=0.021$ for men and $0.022$ for women. Substituting in (\ref{eq:rule}) gives the calibrated rule of thumb
\begin{equation}
  \Dlim \approx 0.71\,\Lth + c\,H , \qquad n_\mathrm{limit}=\Dlim/L_\mathrm{foot}, \label{eq:calib}
\end{equation}
where $n_\mathrm{limit}$ is the same limit expressed in the person's own foot lengths. Equation~(\ref{eq:calib}) is thus not derived analytically but fitted to results of the full model. For the male reference model, $0.71\times86.1+0.021\times181=65.0$~cm. Equation~(\ref{eq:rule}) explains why ankle, knee and neck angles hardly matter: the limit is set by body dimensions, not by the choice of joint angles.

\subsection{Limits in centimetres and in foot lengths}
\begin{figure}[t]
\centering\includegraphics[width=\textwidth]{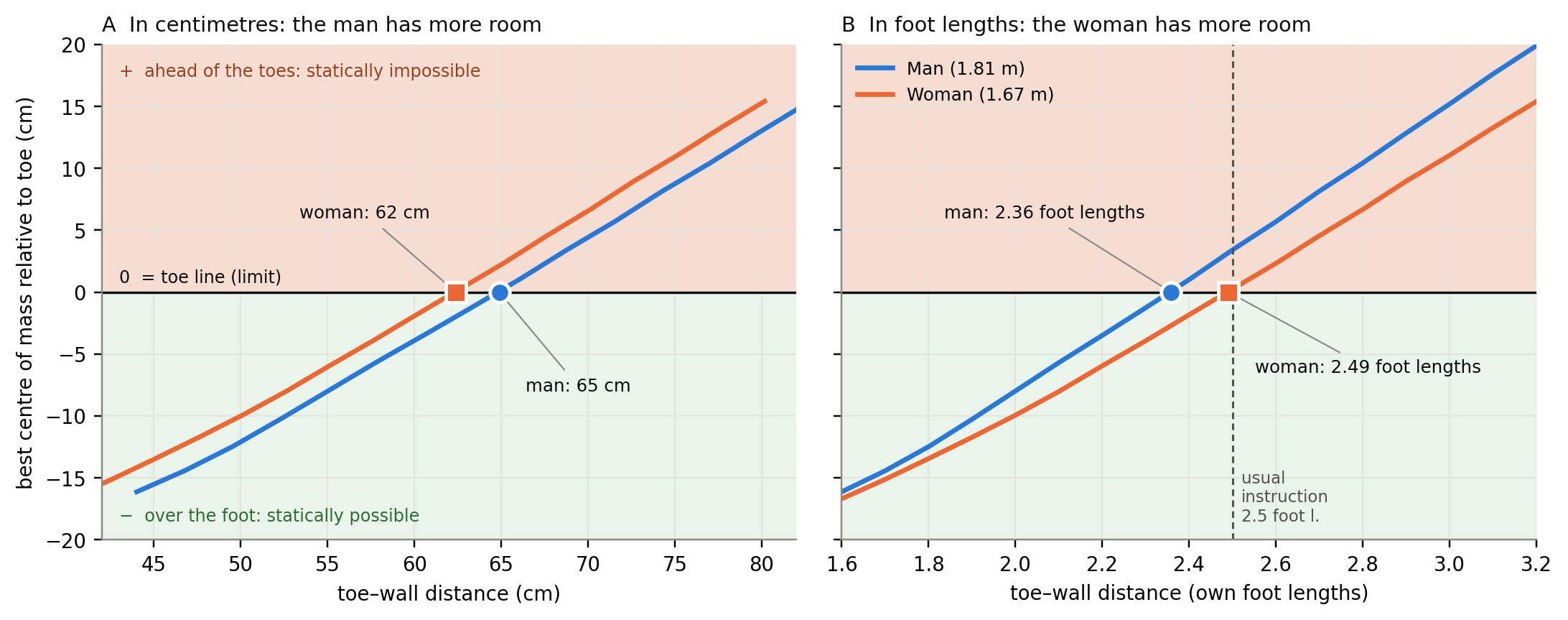}
\caption{Best attainable centre of mass relative to the tip of the toes as a function of the toe--wall distance, in centimetres (A) and in own foot lengths (B), segment proportions of De Leva. Above zero a static rise is impossible; the markers show the geometric limit of each reference model.}
\label{fig:curves}
\end{figure}

Figure~\ref{fig:curves} shows the best attainable centre of mass as a function of the toe--wall distance $D$. In centimetres the man has \emph{more} room than the woman (65 versus 62~cm) because his trunk and head are longer. Only when $D$ is measured in the person's own foot lengths does the woman come out ahead (2.49 versus 2.36\fl). At two foot lengths both succeed; at three, neither does.

The geometric limit assumes that the calf muscles can keep the whole body balanced with the centre of mass above the toe tip. The required ankle torque is about 180~N\,m for the man and 145~N\,m for the woman, or 90 and 73~N\,m per leg. This is about 60\% of the maximum isometric plantar-flexion torque of young adults (18--25 years), which for these body masses and statures is about 140~N\,m per leg for the man and 120~N\,m for the woman~\cite{anderson}, and it has to be held steadily while balancing on the very edge of the support. A practical limit, with the ball of the foot (a quarter foot length behind the toe tip) as the edge of support, lies at 2.05 and 2.19\fl. This is about 0.3 rather than 0.25\fl{} below the geometric limit: closer to the wall the hip cannot move back by the same amount, so the trunk tilts up slightly and the centre of mass moves back only about 0.8~cm per cm of toe--wall distance. The hip extensors need a similar torque to raise trunk, head and chair, about 180 and 150~N\,m in total, but this is only about 35\% of their maximum of roughly 240 and 215~N\,m per leg~\cite{anderson}. The problem is balance, not strength; the advice to ``get your weight into your heels''~\cite{athletico} points the same way.

\subsection{Sensitivity}\label{sec:sens}
Table~\ref{tab:sens} summarises how the geometric limit responds to the assumptions. The first row, the \emph{baseline}, is the model of section~\ref{sec:model} against which all variations are compared: De Leva proportions, reference stature and mass, foot length 15.2\% (man) or 15.0\% (woman) of stature, bare feet, a 4~kg chair held under the abdomen, and the joint ranges of section~\ref{sec:search}. Each other row changes one assumption, for both reference models at once. Chair mass, ankle, knee and neck ranges, hip flexibility and the position of the ankle joint (not listed separately) change the limit by at most 0.03\fl. Holding the chair tightly against the chest costs 0.06. In the alternative segment scaling, trunk, head and arms are scaled with stature alone and the legs make up the remainder, instead of one common factor for all segments. A $\pm3\%$ change in trunk-plus-head length and a $\pm5\%$ change in foot length matter most; stature itself does not, because everything scales with it. The sex difference persists in every variant.

\begin{table}[t]
\caption{Geometric limit, in the person's own foot lengths, when one assumption of the baseline model is changed for both reference models. Baseline: segment proportions of De Leva, reference stature and body mass, bare feet, 4~kg chair held under the abdomen.}
\label{tab:sens}
\centering
\begin{tabular}{lccc}
\toprule
Variation & Man & Woman & Difference \\
\midrule
Baseline (De Leva proportions) & 2.36 & 2.49 & 0.13 \\
Stature $\pm5\%$ & 2.35--2.36 & 2.48--2.49 & 0.13 \\
Foot length $+5\%$ / $-5\%$ & 2.25 / 2.47 & 2.38 / 2.61 & 0.12--0.13 \\
Chair 0~kg / 7~kg & 2.35 / 2.36 & 2.49 / 2.49 & 0.13--0.14 \\
Chair held against chest & 2.30 & 2.43 & 0.13 \\
Trunk + head $+3\%$ / $-3\%$ & 2.42 / 2.29 & 2.56 / 2.42 & 0.13--0.14 \\
Ankle, knee, neck, hip ranges & 2.34--2.36 & 2.46--2.50 & 0.12--0.15 \\
Alternative segment scaling & 2.41 & 2.49 & 0.08 \\
US Army survey proportions, all & 2.37 & 2.40 & 0.03 \\
US Army survey proportions, European ancestry & 2.40 & 2.45 & 0.05 \\
\bottomrule
\end{tabular}
\end{table}

\subsection{A reality check with 6068 soldiers}\label{sec:ansur}
De Leva's parameters are based on 100 men but only 15 women~\cite{deleva}. We therefore replaced the two ratios that matter most, trunk-plus-head length and foot length relative to stature, by those of the 2012 US Army anthropometric survey ANSUR~II, 4082 men and 1986 women with a mean age of 30 years~\cite{ansur}. ANSUR~II contains external body dimensions and body mass but no segment masses; the mass distribution is therefore still taken from De Leva. As trunk-plus-head length we use stature minus trochanterion height; the hip joint centre lies about 3~mm above the trochanterion~\cite{deleva}. In ANSUR~II the trunk plus head is \emph{not} relatively longer in women: 48.1\% of stature against 48.7\% in men (48.6 and 48.9\% in the subgroup of European ancestry). Women's feet are relatively smaller, 15.1 against 15.4\% of stature. Running the full model with these proportions gives limits of 2.37 and 2.40\fl{} (all participants) and 2.40 and 2.45\fl{} (European ancestry). The sex difference drops to 0.03--0.05\fl, and what remains of it comes from relative foot size. The popular foot explanation is thus not entirely wrong, but it concerns the foot relative to the trunk, and the effect is small.

For readers who prefer to think in shoe sizes: one European size corresponds to 2/3~cm of last length~\cite{bergzeit}, so pacing 2.5 feet moves the start position about 1.7~cm per size. The De Leva difference of 0.13\fl{} corresponds to about two shoe sizes, the ANSUR~II difference to half to one size. The standard deviation of foot length within each sex in ANSUR~II is 1.2--1.3~cm, about two sizes.

\subsection{Breasts, body mass, age and shoes}\label{sec:shoes}
\begin{figure}[t]
\centering\includegraphics[width=0.9\textwidth]{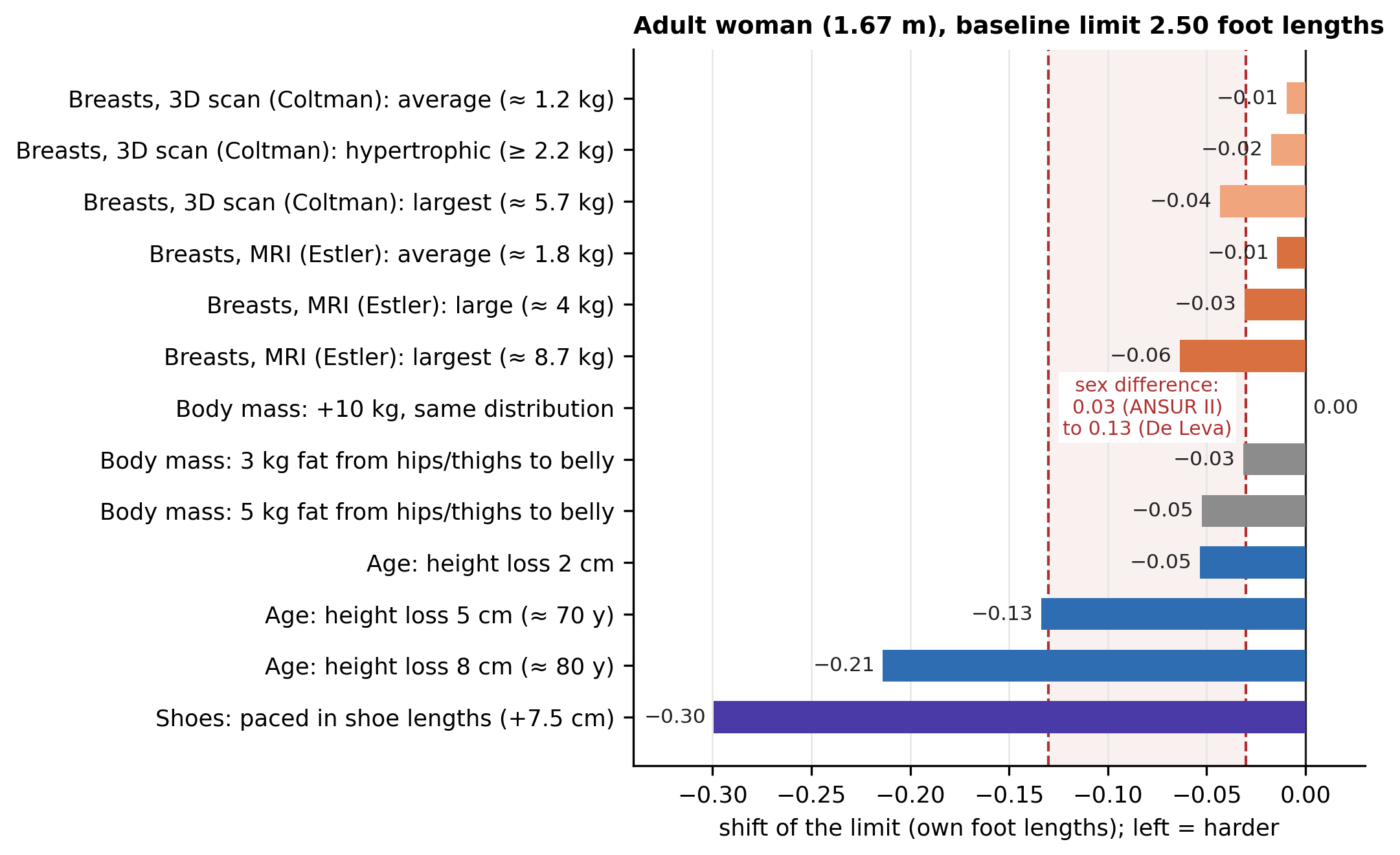}
\caption{Shift of the geometric limit for the female reference model, computed with the simplified fixed best posture (legs vertical, trunk horizontal, head in line; section~\ref{sec:rule}). Breast masses from 3D scanning (Coltman \emph{et al}) and magnetic resonance imaging (Estler \emph{et al}); loss of stature with age from Sorkin \emph{et al}. The shaded band shows the size of the entire sex difference, from 0.03 (US Army survey proportions) to 0.13 foot lengths (De Leva proportions).}
\label{fig:factors}
\end{figure}

The effects in this section were computed with the simplified fixed best posture of section~\ref{sec:rule} (legs vertical, trunk horizontal, head in line), as used for the rule of thumb. A question that is often raised is whether a larger bosom helps. Of course it does not. Breast volumes measured by 3D scanning in the prone position~\cite{coltman} and by magnetic resonance imaging~\cite{estler} correspond to 1.2--1.8~kg for an average and up to 8.7~kg for the largest pair. In the bent posture this mass hangs only about 15~cm ahead of the combined centre of mass and shifts the limit by at most 1.6~cm, in the unfavourable direction (figure~\ref{fig:factors}). Adding body mass with the same distribution changes essentially nothing; moving 3--5~kg of fat from hips to belly costs 0.03--0.05\fl. Age matters more: the loss of stature of about 5~cm in women and 3~cm in men by age 70~\cite{sorkin}, mostly in the spine and hence in $\Lth$, lowers the limit by 0.13 and 0.07\fl.

Shoes matter most of all. A shoe is roughly 2.5~cm longer than the foot, so pacing 2.5 \emph{shoe} lengths places the toes about 7.5~cm further from the wall, which is 0.27--0.30\fl{} and more than twice the largest sex difference we found. With shoes, neither average adult passes the test statically. Because the extra length is about the same for everyone, shoes penalise small feet relatively more and slightly reduce the sex difference.

\subsection{Is the chair challenge a sex test?}
\begin{figure}[t]
\centering\includegraphics[width=0.9\textwidth]{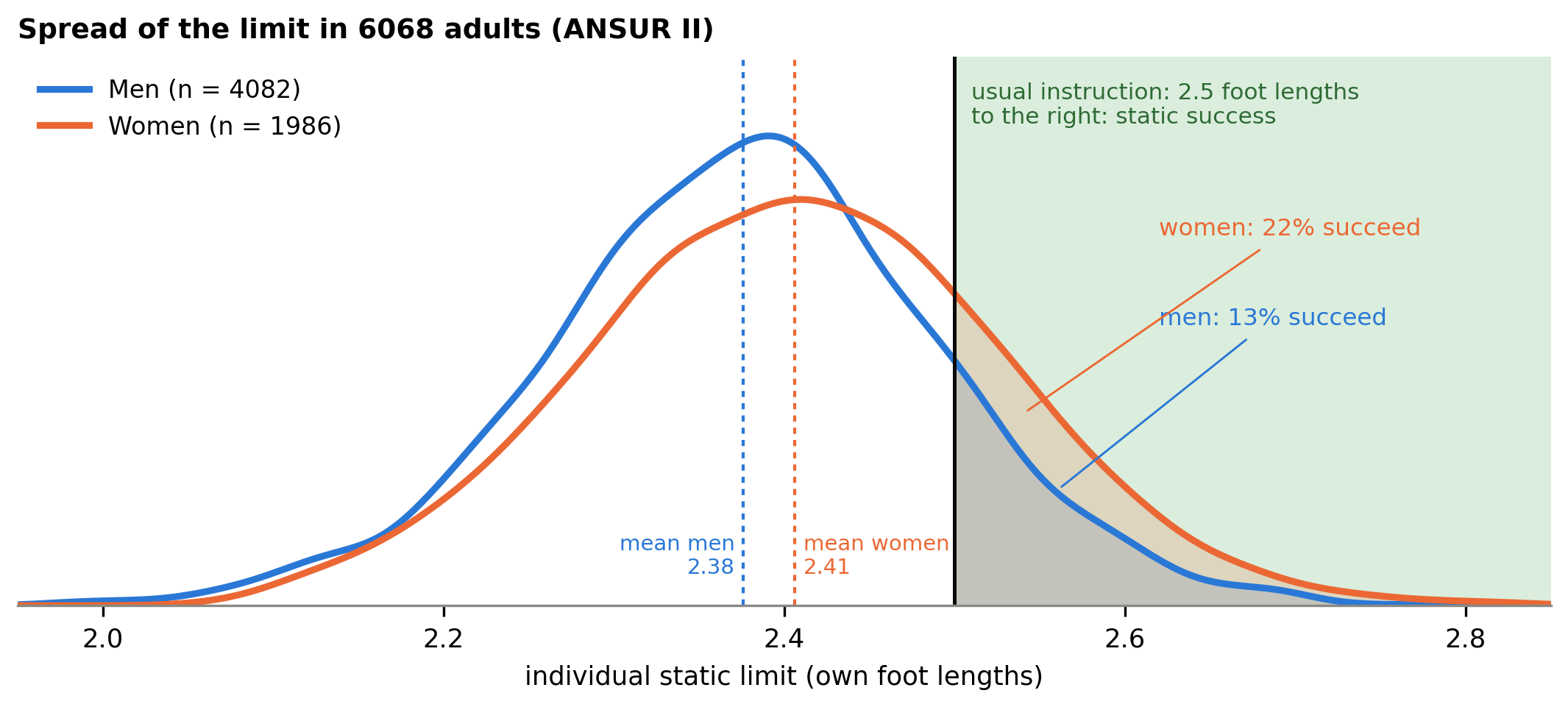}
\caption{Distribution of the individual static limit, in own foot lengths, for the 4082 men and 1986 women of the 2012 US Army anthropometric survey. Dotted lines: means. Participants to the right of the black line (2.5 foot lengths) pass statically.}
\label{fig:spread}
\end{figure}

To judge the trick as a sex test one needs the spread between individuals, not only the averages. Using the calibrated rule of thumb (\ref{eq:calib}) with each ANSUR~II participant's own stature, trochanterion height and foot length, we computed an individual static limit for all 6068 participants (figure~\ref{fig:spread}). The spread within each sex (standard deviation about 0.12\fl, or 3~cm) is several times larger than the difference between the sexes. At the customary 2.5\fl, 13\% of men and 22\% of women would pass statically (16\% and 33\% in the subgroup of European ancestry; 10\% and 46\% with De Leva's proportions). For equal numbers of men and women, predicting ``woman'' after a success and ``man'' after a failure is right in 54\% of cases (58\% and 68\% for the other data sets); even the best possible toe--wall distance raises this only to 56--72\%. A fair coin scores 50\%. Stature or foot length alone, measured with a tape and compared with the best threshold, identifies sex correctly in 83\% of cases. At three foot lengths, a distance also found in the instructions~\cite{kym,cbsny}, not a single one of the 6068 participants passes statically; the largest individual limit is 2.82\fl.

\section{Discussion}
\subsection{Why do so many videos show women succeeding?}
If most people should fail statically at 2.5\fl, why do so many women succeed on camera? We speculate that there are three reasons. First, momentum. Our analysis so far is static, but a real attempt need not be: as long as the head touches the wall, the body forms a strut between floor and wall, and a brisk extension of hips and back pushes the head against the wall and the body backwards; rule R3 allows it. Model the body plus chair as a point mass $m$ at height $h$ above the floor that tips about the toe tip, with its centre of mass a small horizontal distance $x$ ahead of the toes. Linearised, $\ddot x=(g/h)\,x$, so that $\dot x^2-(g/h)x^2$ is conserved. Starting from $x=\Delta$, the centre of mass reaches the toe line only if the initial backward velocity satisfies
\begin{equation}
  v_0 \geq \Delta\sqrt{g/h}. \label{eq:pendulum}
\end{equation}
For $\Delta=5$~cm, $h=0.80$~m and $m=88$~kg (man plus chair) this gives $v_0\approx0.18$~m\,s$^{-1}$ and an impulse $mv_0\approx15$~N\,s, for instance 100~N for 0.15~s. Second, the distance is often paced loosely, sometimes in shoes. Third, and perhaps most importantly, videos in which the expected pattern appears are more likely to be recorded, kept and shared, a selection effect akin to publication bias. Without experiments of our own, these effects could be separated by a systematic analysis of existing videos, in which two independent raters record for each attempt the toe--wall distance (estimated in foot lengths from a side view), whether shoes were worn, whether the person rose slowly or with a jerk, and the apparent sex of the person.

\subsection{Myths revisited}
The ``lower centre of mass'' explanation fails for an instructive reason: what matters is not the height of the centre of mass when standing, but its horizontal distance $\dcom$ from the hip in the bent posture, and that is about 12\% of stature in both sexes (11.7\% for the male and 12.0\% for the female reference model, section~\ref{sec:rule}; if anything to the woman's disadvantage). The ``bigger feet'' explanation fails in absolute terms, since the man has more room in centimetres, but survives in relative form: in ANSUR~II, women's slightly smaller feet relative to their stature are the whole source of their small advantage.

\subsection{The chair challenge in the classroom}\label{sec:teaching}
The problem fits a first-year undergraduate course in mechanics, and parts of it suit a computational or biomechanics project. It touches several ideas that students often meet only separately.
\begin{itemize}
\item \emph{Centre of mass of an articulated body.} The combined centre of mass is a weighted sum of segment centres that moves when the joints move. The question ``where can it be?'' replaces ``where is it?''; the constraints (feet fixed, head on the wall, hands on the chair) turn this into a small optimisation problem.
\item \emph{Static equilibrium and the base of support.} A slow rise is possible only if the vertical through the centre of mass passes through the base of support. The distinction between the geometric limit (toe tip) and the practical limit (ball of the foot, finite ankle torque) shows how a mathematical boundary becomes a physiological one.
\item \emph{Scaling and units.} Because all lengths scale with stature, the limit expressed in own foot lengths is almost independent of stature, while the limit in centimetres is not. Whether men or women ``have more room'' depends on the unit, which is a striking example of why a dimensionless formulation must be chosen with care.
\item \emph{Order-of-magnitude dynamics.} Equation~(\ref{eq:pendulum}) is a two-line estimate with the linearised inverted pendulum that shows why a small push against the wall suffices.
\item \emph{Means versus distributions.} A real difference between the two reference models does not make a good test for individuals when the spread within each group is several times larger. Comparing the hit rate with that of a coin and of a tape measure is a direct way to make this concrete.
\end{itemize}
Students can measure their own stature, trochanterion height (the top of the bony prominence at the side of the hip) and bare foot length, predict their limit with (\ref{eq:calib}), and test it with a light chair, barefoot, at toe--wall distances in steps of a quarter foot length. Comparing slow and brisk attempts shows the role of momentum directly. Because a failed attempt ends with the head against the wall, the risk is small, but a light chair and a non-slip floor are advisable.

\subsection{Limitations}
The model is two-dimensional with a rigid trunk; a curved spine would enlarge the attainable set somewhat. Mass distribution is taken from De Leva for all individuals. The proportions are those of Russian students and US soldiers; other populations, for instance those in the DINED database~\cite{dined}, were not examined. The individual limits rely on the calibrated rule of thumb rather than the full model (see section~\ref{sec:rule}). Finally, the model says nothing about how people actually perform the trick; it only states what is statically possible.

\section{Conclusion}
We modelled the body as a planar chain of rigid segments and searched all postures allowed by the rules of the chair challenge. The human centre of mass is not a fixed point, but with the head against a wall and a chair in the hands only a small region remains attainable. Whether one can stand up slowly depends on the most rearward point of that region, which is set mainly by the length of trunk plus head relative to foot length, not by the height of the centre of mass. In centimetres the male reference model has more room than the female one; only in own foot lengths does the female one come out ahead. With De Leva's proportions the female advantage is 0.13 foot lengths; with the proportions of 6068 soldiers it is 0.03--0.05, half to one shoe size. At the customary distance most people of either sex fail unless they use momentum, and as a test of sex the chair challenge is barely better than a fair coin (54\% against 50\% for the full survey). Shoes, which move the toes about 7.5~cm further from the wall when the distance is paced in shoe lengths, and the loss of stature with age matter more than sex. The main conclusion is that the chair challenge is an excellent classroom demonstration of static equilibrium and a lousy sex predictor. Archie Bunker had a point.

\ack{We thank the anonymous chair for its patience. No chairs, men or women were harmed in this study. During the preparation of this manuscript, the authors used Claude (Anthropic) to support drafting and editing of the text, checking of calculations and references, and formatting for the journal. The authors reviewed, validated and critically assessed all AI-assisted outputs and take full responsibility for the content of the manuscript.}

\section*{Conflict of interest}
The authors declare no competing interests.

\data{The ANSUR~II data are publicly available~\cite{ansur}. The model code (Python) will be deposited in a public repository upon acceptance.}

\end{document}